\documentclass[12pt]{article}

\renewcommand{\thefootnote}{\fnsymbol{footnote}}

\usepackage{amsbsy,amssymb,latexsym,amsfonts,amsmath}
\usepackage{mathrsfs}
\usepackage{graphicx}
\usepackage{bm}
\usepackage{here}

\numberwithin{equation}{section}
\newcommand{\bel}[1]{\begin{equation}\label{#1}}                     
\newcommand{\bal}[1]{\begin{eqnarray}\label{#1}}                     
\newcommand{\be}{\begin{equation}}
\newcommand{\ee}{\end{equation}}
\newcommand{\im}{\mathrm{i}}
\newcommand{\ex}{\mathrm{e}}
\newcommand{\de}{\mathrm{d}}

\newcommand{\qq}{\qquad}

\begin{document}
%
%
\begin{titlepage}
\begin{flushright}
\normalsize
~~~~
OCU-PHYS 465\\
10 May 2017 \\
\end{flushright}

\vspace{15pt}

\begin{center}
{\LARGE Elliptic algebra, Frenkel-Kac construction} \\
\vspace{10pt}
{\LARGE and root of unity limit}
\end{center}

\vspace{23pt}

\begin{center}
{ H. Itoyama$^{a, b}$\footnote{e-mail: itoyama@sci.osaka-cu.ac.jp},
T. Oota$^b$\footnote{e-mail: toota@sci.osaka-cu.ac.jp}
  and  R. Yoshioka$^b$\footnote{e-mail: yoshioka@sci.osaka-cu.ac.jp}  }\\
%
\vspace{18pt}
%

$^a$\it Department of Mathematics and Physics, Graduate School of Science\\
Osaka City University\\
\vspace{5pt}

$^b$\it Osaka City University Advanced Mathematical Institute (OCAMI)

\vspace{5pt}

3-3-138, Sugimoto, Sumiyoshi-ku, Osaka, 558-8585, Japan \\

\end{center}
%
\vspace{20pt}
\begin{center}
Abstract\\
\end{center}
We argue that the level-$1$ elliptic algebra $U_{q,p}(\widehat{\mathfrak{g}})$
is a dynamical symmetry realized as a part of 2d/5d correspondence where the Drinfeld currents are the screening currents to the $q$-Virasoro/W block in the 2d side. For the case of $U_{q,p}(\widehat{\mathfrak{sl}}(2))$,
the level-$1$ module has a realization by an elliptic version of the Frenkel-Kac construction.
The module admits the action of the deformed Virasoro algebra.
In a $r$-th root of unity limit of $p$ with $q^2 \rightarrow 1$, the $\mathbb{Z}_r$-parafermions
and a free boson appear and the value of the central charge that we obtain agrees with 
that of the 2d coset CFT with para-Virasoro symmetry, which corresponds to 
the 4d $\mathcal{N}=2$ $SU(2)$ gauge theory on $\mathbb{R}^4/\mathbb{Z}_r$.


\vfill

\end{titlepage}

\renewcommand{\thefootnote}{\arabic{footnote}}
\setcounter{footnote}{0}

\section{Introduction}

Since the proposal of AGT(W) relation \cite{AGT0906,wyl0907},
2d/4d correspondence and its generalizations have been intensively studied.
Originally, it was the correspondence between the Nekrasov partition function of
4d $\mathcal{N}=2$ supersymmetric $SU(N)$
gauge theory and the conformal block of 2d CFT with the $W_N$ symmetry.
One of the generalizations is to consider the 4d $SU(N)$ gauge theory on 
$\mathbb{R}^4/\mathbb{Z}_r$ \cite{wyl1109,AT1110}. 
For works in this direction, see for example \cite{BFL1011,BF1105,BBB1106,BMT,AT1110,I1110,BBFLT1111,BBT1211,ABT1306,IOY2,IOY1408,BPSS1312}.
The corresponding CFT is described by a coset
\be
\frac{\widehat{\mathfrak{sl}}(N)_r \oplus \widehat{\mathfrak{sl}}(N)_{\kappa}}
{\widehat{\mathfrak{sl}}(N)_{r+\kappa}},
\ee
posessing the $r$-th ``para-$W_N$ symmetry'' \cite{BF1105,NT1106}. Here $\kappa$ is a parameter
related to the $\Omega$-background parameters $\epsilon_{1,2}$.

Another generalization is a 5d lift (or K-theoretic lift), i.e., to
consider the 5d $\mathcal{N}=1$ $SU(N)$ gauge theories on 
$\mathbb{R}^4 \times S^1$ \cite{AY0910,
NPP1303,Orl1310,tan1301,AHKS1309,yos1512,KP1512}.
(For more general 2d/6d correspondence, see for example \cite{sai1301,Tan1309,IKY1511,nie1511,KP1608}).
The corresponding 2d theories are no longer conformally invariant but
are invariant under the deformed Virasoro \cite{LP9412,FR95,SKAO9507} ($N=2$) or deformed $W_N$ symmetry ($N \geq 3$) \cite{FF9508,AKOS9508}.
The deformed vertex operators \cite{AKMOS9604,kad,JS97,IOY1602} 
play important roles in constructing
$q$-deformed conformal/W block.
The $q$-Virasoro/W algebras themselves are, however,  not sufficient to
determine the $q$-vertex operators. This comes
from their lack of the coalgebra structure---in particular,  the coproduct.
To construct the deformed vertex operators, one can take an approach
that utilizes algebras having the coproduct and that is closely connected with 
the $q$-Viraroso/W algebras. There are at least two
such algebras: the Ding-Iohara-Miki (DIM) algebras \cite{DI9608a,mik07}
and the elliptic algebra
$U_{q,p}(\widehat{\mathfrak{g}})$ \cite{kon9709,JKOS9802,KK0210,KK0307,KKW0504,kon0802,kon0803,kon1603}. 
Here $\widehat{\mathfrak{g}}$
is an untwisted affine Lie algebra.
They are different kinds of extension
of the quantum group $U_q(\widehat{\mathfrak{g}})$.
For researches based on the DIM algebras, see, for example, 
\cite{AFHKSY,KMZ1308,spo1409,IKY1511,OAF1512,AKMMMMOZ1604,AKMMMOZ1608,MMZ1603a,BFMZZ1606}.

In this paper, we take the second approach to make exploit the elliptic algebra $U_{q,p}(\widehat{\mathfrak{g}})$. One important property of this elliptic algebra with regard to the 2d/5d
correspondence is that the Drinfeld currents act as the screening currents on the $q$-Virasoro/W block
in the 2d side.
The elliptic algebras are constructed based on the elliptic solutions to the Yang-Baxter equations.
There are two class of elliptic solutions. One is related to the eight-vertex model (XYZ model) \cite{bax72,bel81}. 
The other is related to the face-type integrable lattice models (ABF \cite{ABF84} or RSOS models).
The corresponding elliptic algebras are called vertex-type and face-type respectively.
The Sklyanin algebra \cite{skl82} and $\mathcal{A}_{q,p}(\widehat{\mathfrak{sl}}(N))$
\cite{FIJKMY9403,JKOS9712} are vertex-type elliptic algebras.

The face-type elliptic solutions obey the dynamical Yang-Baxter equation  (or the Gervais-Neveu-Felder equation) \cite{GN84,fel9407}. The dynamical Yang-Baxter equation
\be
R_{12}(\lambda + h_3) R_{13}(\lambda) R_{23}(\lambda+h_1)
= R_{23}(\lambda) R_{13}(\lambda+h_2) R_{12}(\lambda)
\ee
is equivalent to the star-triangle equation in solvable lattice models \cite{fel9412}.
Here $\lambda$ is a dynamical parameter and $h$ is an element of the Cartan subalgebra.
This equation first appeared in the study of the monodromy properties
of the conformal blocks of the Liouville field theory \cite{GN84}. (See also
\cite{bab91,BBB96,ABB9505}).

Based on the face-type elliptic solutions, various elliptic quantum groups have been introduced. Some
of them are 
$E_{\tau, \eta}(\mathfrak{sl}(2))$ \cite{fel9407,fel9412}, $\mathcal{B}_{q,\lambda}(\widehat{\mathfrak{g}})$ \cite{JKOS9712} and $U_{q,p}(\widehat{\mathfrak{g}})$.
The latter two, $\mathcal{B}_{q,\lambda}(\widehat{\mathfrak{g}})$
and $U_{q,p}(\widehat{\mathfrak{g}})$,  
are face-type algebras closely related each other. They are quite similar, but have
different Hopf algebra-like structures.
$\mathcal{B}_{q, \lambda}(\widehat{\mathfrak{g}})$ is a quasi-Hopf algebra \cite{dri89}
whose coproduct is not coassociative, while $U_{q,p}(\widehat{\mathfrak{g}})$
is a $H$-Hopf algebroid \cite{EV9708} whose coproduct is coassociative. 
Due to this coassociativity, $U_{q,p}(\widehat{\mathfrak{g}})$ has
a simpler coalgebra structure than $\mathcal{B}_{q, \lambda}(\widehat{\mathfrak{g}})$ does.

Let us consider the face-type elliptic algebra
$U_{q,p}(\widehat{\mathfrak{g}})$ with level $k$. It is an elliptic deformation of the algebra
of screening charges of the coset CFT \cite{kon9709,JKOS9802}
\be
\frac{\widehat{\mathfrak{g}}_k \oplus \widehat{\mathfrak{g}}_{s-k-2}}
{\widehat{\mathfrak{g}}_{s-2}}.
\ee
This CFT is closely related to the $k$-fusion RSOS model of type $\mathfrak{g}$.
For the level $k=1$, $U_{q,p}(\widehat{\mathfrak{g}})$ is closely related to
the deformed $W$ algebra of type $\mathfrak{g}$ \cite{FF9508,AKOS9508}. 
In the CFT limit ($q \rightarrow 1$ limit),
the coset model $\widehat{\mathfrak{g}}_1 \oplus \widehat{\mathfrak{g}}_{s-3}/\widehat{\mathfrak{g}}_{s-2}$ shows the ordinary $W(\mathfrak{g})$-symmetry.

In this paper, we would like to argue that the level-$1$ $U_{q,p}(\widehat{\mathfrak{sl}}(N))$ algebra
is a dynamical symmetry  in its connection with the 5d $\mathcal{N}=1$ $SU(N)$ gauge
theory on $\mathbb{R}^4 \times S^1$. Let us denote the radius of $S^1$ by $R$.
We propose the following dictionary between the deformation parameters and gauge theory parameters:
\bel{dict}
q = \ex^{(1/2)R(\epsilon_1+\epsilon_2)}, \qq
p = \ex^{R \epsilon_2}, \qq
p^* = p q^{-2} = \ex^{-R \epsilon_1}.
\ee
Taking $R \rightarrow 0$ limit is equivalent to a CFT limit of $U_{q,p}(\widehat{\mathfrak{sl}}(N))$
with $q\rightarrow 1$ ($p, p^* \rightarrow 1$). Hence we obtain ordinary 2d/4d correspondence:
4d $SU(N)$ gauge theory on $\mathbb{R}^4$ and 2d CFT with $W_N$ symmetry.

We also propose another CFT limit: a root of unity limit of the parameters
\be
p \rightarrow \omega^{\ell}, \qq
p^*  \rightarrow \omega^{\ell}, \qq q^2 \rightarrow 1,
\ee
where $\omega$ is the primitive $r$-th root of unity and $\ell$ is an integer such that $\omega^{\ell} \neq 1$. If 
the level $1$ elliptic algebra $U_{q,p}(\widehat{\mathfrak{sl}}(N))$ is
the 2d side symmetry of 2d/5d correspondence and deformed blocks
are given by the correlation functions of vertex operators of this algebra, 
this root of unity limit automatically
leads to the  correspondence between the $W$ block of 
the coset CFT $\widehat{\mathfrak{sl}}(N)_r \oplus \widehat{\mathfrak{sl}}(N)_{\kappa}/
\widehat{\mathfrak{sl}}(N)_{r+\kappa}$ and the  Nekrasov instanton partition
function on $\mathbb{R}^4/\mathbb{Z}_r$.

For simplicity, we consider the $N=2$ case: $U_{q,p}(\widehat{\mathfrak{sl}}(2))$ \cite{kon9709}.
Generalization to general $N$ is straightforward (though it may be tedious).

For the level $k=1$  elliptic algebra $U_{q,p}(\widehat{\mathfrak{sl}}(2))$, 
the closely related RSOS model is known as the Andrews-Baxter-Forrester (ABF) face model \cite{ABF84}.
The $q$-deformed Virasoro algebra plays the role of dynamical symmetry
of the ABF model \cite{LP9412,LP9602,JLMP9607,pug04}.
It has a general level $k$ realization by a deformed $Z$-algebra or parafermions.
As we here only deal with its connection with the deformed Virasoro algebra, 
we consider a simple level-$1$ realization, i.e., an elliptic deformation
of the Frenkel-Kac construction. In the $p \rightarrow 0$ limit, the elliptic algebra
$U_{q,p}(\widehat{\mathfrak{sl}}(2))$ essentially goes to 
the quantum group $U_q(\widehat{\mathfrak{sl}}(2))$, and if we take
further limit $q \rightarrow 1$, it goes to the affine Lie algebra $\mathfrak{sl}(2)_k$.

This paper is organized as follows: In the next section, we review the elliptic algebra 
$U_{q,p}(\widehat{\mathfrak{sl}}(2))$. In section 3, level $1$-modules of
$U_{q,p}(\widehat{\mathfrak{sl}}(2))$ algebra based on an elliptic version of the Frenkel-Kac construction
is explained. In section 4, we discuss the root of unity limit of the level-$1$ $U_{q,p}(\widehat{\mathfrak{sl}}(2))$ algebra. 
In appendix A, we briefly recall the Frenkel-Kac construction of the affine Lie algebra 
$\widehat{\mathfrak{sl}}(2)_1$.

\section{Elliptic algebra $U_{q,p}(\widehat{\mathfrak{sl}}(2))$}

In this section, we review the elliptic algebra $U_{q,p}(\widehat{\mathfrak{sl}}(2))$ \cite{kon9709}.
The face-type elliptic algebra $U_{q,p}(\widehat{\mathfrak{sl}}(2))$
is an elliptic deformation of the affine Lie algebra $\widehat{\mathfrak{sl}}(2)$.
If we take the deformation parameters $p \rightarrow 0$ and $q \rightarrow 1$, then $U_{q,p}(\widehat{\mathfrak{sl}}(2))$ goes to the $\widehat{\mathfrak{sl}}(2)$ current algebra (and a Heisenberg algebra).
We essentially follow the convention of \cite{FKO1404}. 

Let $q$ and $p$ be two parameters.
The elliptic algebra $U_{q,p}(\widehat{\mathfrak{sl}}(2))$
is a unital associative algebra generated by the following elements
\bel{genellip}
P, h, e_m, f_m, \alpha_n, K^{\pm}, q^{\pm (1/2) k}, d, \qq
(m \in \mathbb{Z}, n \in \mathbb{Z}\setminus\{0\}).
\ee
$K^{\pm}$ are invertible, $k$ is a central element and $d$ is a grading operator
\be
[d, e_m] = m \, e_m, \qq
[ d, f_m] =m \, f_m, \qq
[ d, \alpha_n] = n \, \alpha_n,
\qq
[ d, P ] = [ d, h ] = [ d, K^{\pm} ] =0.
\ee
The eigenvalue of $k$ on a $U_{q,p}(\widehat{\mathfrak{sl}}(2))$-module is called the level of the module.

It is convenient to introduce the elliptic currents:
\be
e(z) = \sum_{m \in \mathbb{Z}} e_m z^{-m-1}, \qq
f(z) = \sum_{m \in \mathbb{Z}} f_m z^{-m-1},
\ee
\bel{psip}
\psi^+(q^{-k/2} z) = K^+ 
\exp\left( - (q-q^{-1}) \sum_{m>0} \frac{\alpha_{-m}}{1-p^m} z^m \right)
\exp\left( (q-q^{-1}) \sum_{m>0} \frac{p^m \alpha_m}{1-p^m} z^{-m} \right),
\ee
\bel{psim}
\psi^-(q^{k/2}z) = K^- \exp\left( - (q-q^{-1}) \sum_{m>0} \frac{p^m \alpha_{-m}}{1-p^m} z^m
\right)
\exp\left( (q-q^{-1}) \sum_{m>0} \frac{\alpha_m}{1-p^m} z^{-m} \right).
\ee
Then the remaining defining relations are given by
\be
[P, h ] = 0,
\qq
[ P, e(z) ] = - 2 \, e(z), \qq
[ h, e(z) ] = 2 \, e(z),
\ee
\be
[ P, f(z) ] = 0, \qq
[ h, f(z) ] = - 2 \, f(z),
\qq
[ P, \alpha_n] = 0, \qq [h, \alpha_n ] = 0,
\ee
\be
[ P, K^{\pm} ] = -2 K^{\pm}, \qq [h, K^{\pm}] = 0,
\ee
\be
K^{\pm} e(z) = q^{\mp 2} e(z) K^{\pm}, \qq
K^{\pm} f(z) = q^{\pm 2} f(z) K^{\pm},
\ee
\bel{comalpha}
[ \alpha_m, \alpha_n] = \delta_{m+n,0} \frac{[2m]_q [ km]_q}{m} \frac{1-p^{|m|}}{1-p^{*|m|}} q^{-k|m|},
\ee
\be
[ \alpha_m, e(z) ] = \frac{[2m]_q}{m} \frac{1-p^{|m|}}{1-p^{*|m|}} q^{-k|m|} z^m e(z),
\qq
[ \alpha_m, f(z) ] = - \frac{[2 m]_q}{m} z^m f(z),
\ee
\bel{Defee}
z_1 \frac{(q^2 z_2/z_1; p^*)_{\infty}}{(p^* q^{-2} z_2/z_1; p^*)_{\infty}}
e(z_1) e(z_2) = - z_2 \frac{(q^2 z_1/z_2; p^*)_{\infty}}{(p^* q^{-2} z_1/z_2; p^*)_{\infty}}
e(z_2) e(z_1),
\ee
\bel{Defff}
z_1 \frac{(q^{-2} z_2/z_1; p)_{\infty}}{(p q^{2} z_2/z_1; p)_{\infty}}
f(z_1) f(z_2) = - z_2 \frac{(q^{-2} z_1/z_2; p)_{\infty}}{(p q^{2} z_1/z_2; p)_{\infty}}
f(z_2) f(z_1),
\ee
\bel{Defef}
[ e(z_1), f(z_2)] = \frac{1}{(q-q^{-1})z_1 z_2}
\Bigl( \delta(q^{-k} z_1/z_2) \psi^-(q^{k/2} z_2) - \delta(q^k z_1/z_2) \psi^+(q^{-k/2} z_2)
\Bigr).
\ee
Here $p^* = p q^{-2k}$ and
\be
[ x ]_q = \frac{q^x - q^{-x}}{q-q^{-1}}, \qq
( x; \xi)_{\infty} = \prod_{n=0}^{\infty} ( 1 - x \xi^n ), \qq
\delta(z) = \sum_{n \in \mathbb{Z}} z^n.
\ee

\subsection{$p \rightarrow 0$ limit: $U_{q}(\widehat{\mathfrak{sl}}(2))$}

In the $p \rightarrow 0$ limit, the elliptic algebra $U_{q,p}(\widehat{\mathfrak{sl}}(2))$
becomes the direct product of quantum group $U_q(\widehat{\mathfrak{sl}}(2))$
and the algebra $\mathcal{H}$ generated by $\{ P, \ex^{\pm 2 Q} \}$ where $[P, Q] = 1$.
\be
U_{q,p}(\widehat{\mathfrak{sl}}(2)) \longrightarrow  U_q(\widehat{\mathfrak{sl}}(2))
\otimes \mathcal{H}, \qq
[ U_q(\widehat{\mathfrak{sl}}(2)), \mathcal{H} ] = 0,
\ee
\be
P \rightarrow P, \qq
h \rightarrow h, \qq
q^{\pm (1/2) k} \rightarrow q^{\pm (1/2) k}, \qq
d \rightarrow \tilde{d},
\ee
\be
e(z) \rightarrow x^+(z) \ex^{-2Q}, 
\qq
f(z) \rightarrow x^-(z),
\qq
\alpha_n \rightarrow \widetilde{\alpha}_n, 
\qq
K^{\pm} \rightarrow q^{\mp h} \ex^{-2Q},
\ee
\be
\psi^+(q^{-k/2} z) \rightarrow  \varphi(q^{-k/2} z) \ex^{-2Q}, \qq
\psi^-(q^{k/2}z) \rightarrow \psi(q^{k/2} z) \ex^{-2Q},
\ee
where 
\be
\varphi(q^{-k/2}z) = q^{-h} \exp\left( - (q-q^{-1}) \sum_{m>0} \widetilde{\alpha}_{-m} z^m \right),
\ee
\be
\psi(q^{k/2}z) = q^{h} \exp\left( (q-q^{-1}) \sum_{m>0} \widetilde{\alpha}_m z^{-m} \right).
\ee
Here $\widetilde{\alpha}_n$ satisfies the following commutation relations:
\be
[ \widetilde{\alpha}_m, \widetilde{\alpha}_n] = \delta_{m+n,0} \frac{[2m]_q [ km]_q}{m} q^{-k|m|}.
\ee

The mode expansion of the Drinfeld currents $x^{\pm}(z)$ is given by
\be
x^{\pm}(z) = \sum_{m \in \mathbb{Z}} x^{\pm}_m z^{-m-1}.
\ee
Then, the quantum group $U_q(\widehat{\mathfrak{sl}}(2))$ is generated by 
\be
x^{\pm}_m, \widetilde{\alpha}_n, h, q^{\pm (1/2)k}, \tilde{d},
 \qq
(m \in \mathbb{Z}; n \in \mathbb{Z} \setminus \{0\}).
\ee
The defining relations are given as follows: $k$ is a central element,
\be
[ \tilde{d}, x^{\pm}_m ] = m x^{\pm}_m, \qq
[ \tilde{d}, \widetilde{\alpha}_n ] = n \widetilde{\alpha}_n, \qq
[ \tilde{d}, h ] = 0,
\ee
\be
[ h, \tilde{\alpha}_n] = 0, \qq
[ h, x_n^{\pm} ] = \pm 2 x_n^{\pm},
\ee
\be
[ \widetilde{\alpha}_m, \widetilde{\alpha}_n ] = \delta_{m+n,0} \frac{[2m]_q [km]_q}{m} q^{-k|m|},
\ee
\be
[ \widetilde{\alpha}_m, x^{+}(z) ] = \frac{[2m]_q}{m} q^{-k|m|} z^m x^{+}(z), \qq
[ \widetilde{\alpha}_m, x^{-}(z) ] = - \frac{[2m]_q}{m} z^m x^-(z),
\ee
\be
z_1(1 -q^{\pm 2} z_2/z_1 ) x^{\pm}(z_1) x^{\pm} (z_2) = 
-z_2( 1 - q^{\pm 2} z_1/z_2) x^{\pm} (z_2) x^{\pm}(z_1),
\ee
\be
[ x^+(z_1), x^-(z_2) ]
= \frac{1}{(q-q^{-1})z_1 z_2} \Bigl( \delta(q^{-k}z_1/z_2) \psi(q^{k/2}z_2)
- \delta(q^k z_1/z_2) \varphi(q^{-k/2} z_2) \Bigr).
\ee

\subsubsection{$q \rightarrow 1$: $\widehat{\mathfrak{sl}}(2)_k$ current algebra}

Furthermore, if we also take the $q \rightarrow 1$ limit, $U_q(\widehat{\mathfrak{sl}}(2))$ goes
to the $\widehat{\mathfrak{sl}}(2)$ current algebra with level $k$:
\be
U_q(\widehat{\mathfrak{sl}}(2)) \rightarrow \widehat{\mathfrak{sl}}(2)_k,
\ee
\be
x^{\pm}(z) \rightarrow J^{\pm}(z), 
\qq
\widetilde{\alpha}_n \rightarrow a_n,
\qq
\frac{ \psi(q^{k/2} z) - \varphi(q^{-k/2} z)}{(q-q^{-1})z}
\rightarrow 2 J^3(z),
\ee
where 
\be
2 J^3(z) =  h z^{-1} + \sum_{m \neq 0} a_m z^{-m-1}.
\ee
Here
\be
[a_m, a_n ] = 2\, k \, m \delta_{m+n,0}.
\ee
The commutation relations for the $\widehat{\mathfrak{sl}}(2)$ currents are given by
\be
[ J^3(z_1), J^{\pm}(z_2) ] = \pm \delta(z_2/z_1) \frac{J^{\pm}(z_2)}{z_2},
\qq
[ J^3(z_1), J^3(z_2) ] = \frac{(k/2)}{z_1 z_2} \delta'(z_2/z_1),
\ee
\be
[J^+(z_1), J^-(z_2)] = \frac{k}{z_1 z_2} \delta'(z_2/z_1)
+\delta(z_2/z_1) \frac{2 J^3(z_2)}{z_2}.
\ee
Here $\delta'(x) = \sum_{n \in \mathbb{Z}} n x^n$.

These commutation relations are equivalent to the following OPE:
\be
J^3(z_1) J^{\pm}(z_2) \sim \frac{\pm J^{\pm}(z_2)}{z_1-z_2},
\qq
J^3(z_1) J^3(z_2) \sim \frac{k/2}{(z_1-z_2)^2},
\ee
\be
J^+(z_1) J^-(z_2) \sim \frac{k}{(z_1-z_2)^2} + \frac{2 J^3(z_2)}{z_1 - z_2}.
\ee

\section{Level $1$ modules of $U_{q,p}(\widehat{\mathfrak{sl}}(2))$}


It is well-known that the level $k=1$ modules of the $\widehat{\mathfrak{sl}}(2)$ current algebra can be
obtained by a free massless chiral boson compactified on a circle with the self-dual radius. 
The Fock space of the compactified boson is decomposed into the two irreducible $\widehat{\mathfrak{sl}}(2)_1$
modules with highest weights $\Lambda_0$ and $\Lambda_1$.
It is the so-called
the Frenkel-Kac construction \cite{FK80} (see Appendix \ref{FKC} for a brief review).

In this section, the elliptic analog of the Frenkel-Kac construction is explained.

Let us introduce elliptic bosons by
\be
\Phi(z) = 2 Q_h + h \log z - \sum_{m \neq 0} \frac{\alpha_m}{[m]_q} z^{-m},
\ee
\be
\Phi^{\vee}(z) = 2 Q_h + h \log z - \sum_{m \neq 0} \frac{\alpha_m}{[m]_q}
\frac{(1-p^{*|m|})}{(1-p^{|m|})} q^{|m|} z^{-m},
\ee
where the modes obey the following commutation relations
\be
[ \alpha_m, \alpha_n] = \delta_{m+n,0} \frac{[2m]_q [m]_q}{m} 
\frac{(1-p^{|m|})}{(1-p^{*|m|})} q^{-|m|},
\qq [ h, Q_h ]=1, \qq (m,n \neq 0).
\ee
Note that the commutation relations among the non-zero modes $\alpha_n$
are the $k=1$ case of \eqref{comalpha}.
We also introduce an additional algebra generated by $\{ P, \ex^{\pm 2 Q} \}$ with $[ P, Q ] = 1$.

We assume that the elliptic bosons are ``compactified on a circle of self-dual radius''. This means that
the eigenvalues of $h$ on the elliptic boson Fock space are integers\footnote{See a remark on the last paragraph of Appendix \ref{FKC}}. 
Hence $Q_h$ can appear only in
the form $\ex^{n Q_h}$ ($n \in \mathbb{Z}$).

Let $|0 \rangle$ be the Fock vacuum characterized by
\be
h \, | 0 \rangle = 0, \qq
P \, | 0 \rangle = 0, \qq
\alpha_{n} | 0 \rangle = 0, \qq
(n>0).
\ee
The Fock space $\mathcal{F}$ of the elliptic bosons and the additional algebra is spanned by the following vectors
\be
\alpha_{-n_k} \alpha_{-n_{k-1}} \dotsm \alpha_{-n_2} 
\alpha_{- n_{1}} \ex^{2 m_1 Q + m_2 Q_h} | 0 \rangle, \qq
(0<n_1 \leq n_2 \leq \dotsm  \leq n_k; m_1, m_2 \in \mathbb{Z}).
\ee

The action of the level $1$  $U_{q,p}(\widehat{\mathfrak{sl}}(2))$ on the Fock space $\mathcal{F}$
is realized by 
\be
e(z) = :\ex^{-2Q} \, \ex^{\Phi(z)}:, \qq
f(z) = :\ex^{- \Phi^{\vee}(z)}:.
\ee
\bel{level1Kpm}
K^{\pm} = \ex^{-2Q} q^{\mp h}.
\ee
Here $p^* = p q^{-2}$. The normal ordering is defined by moving $Q$ and $Q_h$ to the left of
$P$ and $h$, the creation operators $\alpha_{-n}$ ($n>0$)  to the left of the annihilation
operators $\alpha_n$ $(n>0)$.  Hence, 
\be
e(z) = \ex^{-2Q + 2Q_h} z^{h}
\exp\left( \sum_{m>0} \frac{\alpha_{-m}}{[m]_q} z^m \right)
\exp\left( - \sum_{m>0} \frac{\alpha_m}{[m]_q} z^{-m} \right),
\ee
\be
f(z) = \ex^{-2Q_h} z^{-h}
\exp\left( - \sum_{m>0} \frac{\alpha_{-m}}{[m]_q}
\frac{(1-p^{*m})}{(1-p^m)} q^m z^m \right)
\exp\left( \sum_{m>0} \frac{\alpha_m}{[m]_q}
\frac{(1-p^{*m})}{(1-p^m)} q^m z^{-m} \right).
\ee

The currents $\varphi^{\pm}(q^{\mp k/2}z)$ are given by \eqref{psip} and \eqref{psim}
for level $1$ $\alpha_m$ and with $K^{\pm}$ substituted by \eqref{level1Kpm}.
The grading operator is realized as
\be
d = - \frac{1}{4} h^2 - 
\sum_{m>0}
\frac{m^2}{[2m]_q [m]_q} \frac{(1-p^{*m})}{(1-p^{m})} q^m \alpha_{-m} \alpha_m.
\ee
As in the case of the affine Lie algebra, the Fock space $\mathcal{F}$ decomposes into two irreducible level $1$ 
$U_{q,p}(\widehat{\mathfrak{sl}}(2))$-modules according to the eigenvalues of $(-1)^{h}$:
\be
\mathcal{F} = \mathcal{F}_+ \oplus \mathcal{F}_-,
\ee
where
$\mathcal{F}_{\pm} = \{ v \in \mathcal{F}\, | \, (-1)^h v = \pm v \}$.

\subsection{$p \rightarrow 0$ limit}

In the $p \rightarrow 0$ limit, we obtain the free boson representation of level $1$ $U_q(\widehat{\mathfrak{sl}}(2))$
\cite{FJ88}:
\be
e(z) \rightarrow \ex^{-2Q} x^+(z), \qq
f(z) \rightarrow x^{-}(z), \qq \alpha_n \rightarrow \widetilde{\alpha}_n,
\ee
where
\be
x^+(z) = 
\ex^{2Q_h} z^{h} : \exp\left( - \sum_{m \neq 0}
\frac{\widetilde{\alpha}_{m}}{[m]_q}  z^{-m} \right):,
\ee
\be
x^-(z) = \ex^{-2Q_h} z^{-h}
: \exp\left( \sum_{m \neq 0} \frac{q^{|m|} \widetilde{\alpha}_m}{[m]_q}
z^{-m} \right):,
\ee
with
\be
[ \widetilde{\alpha}_m, \widetilde{\alpha}_n ] = \delta_{m+n,0} \frac{[2m]_q[m]_q}{m} q^{-|m|}.
\ee
Note that
\be
\tilde{d} = - \frac{1}{4} h^2 - \sum_{m>0}
\frac{m^2}{[2m]_q[m]_q} q^{m} \widetilde{\alpha}_{-m} \widetilde{\alpha}_m.
\ee

\subsubsection{$q \rightarrow 1$ limit}

Furthermore, if we take $q \rightarrow 1$ limit, we obtain the Frenkel-Kac construction
of the level $1$ $\widehat{\mathfrak{sl}}(2)$ current algebra:
\be
x^{\pm}(z) \rightarrow  J^{\pm}(z), \qq
\widetilde{\alpha}_n \rightarrow a_n,
\ee
where
\be
J^{\pm}(z) = : \ex^{\pm \Phi_0(z)}:
= \ex^{\pm 2Q_h} z^{\pm h} : \exp\left( \mp \sum_{m \neq 0} \frac{a_m}{m} z^{-m} \right):,
\ee
\bel{Phiz}
\Phi_0(z) = 2 Q_h + h \log z - \sum_{m \neq 0} \frac{a_m}{m} z^{-m},
\qq
[ a_m, a_n] = 2 m \delta_{m+n,0}.
\ee
Also, we find
\be
J^3(z) = \frac{1}{2} \partial \Phi_0(z).
\ee
Note that the normalization of the free boson $\Phi(z)$ is chosen as
$\langle \Phi_0(z_1) \Phi_0(z_2) \rangle = 2 \log (z_1 - z_2)$.

In the $q \rightarrow 1$ limit, the grading operator goes to
\be
\tilde{d} \rightarrow - L_0,
\ee
where
\be
L_0 = \frac{1}{4} h^2 + \frac{1}{2} \sum_{m>0} a_{-m} a_m.
\ee
This operator $L_0$ is one of the Virasoro generators with the central charge $c=1$:
\be
T(z) = \frac{1}{4} : \bigl(\partial \Phi_0(z) \bigr)^2: = \sum_{n \in \mathbb{Z}} L_n z^{-n-2}.
\ee

\subsection{$q$-Virasoro algebra}

It is known that on the level-$1$ $U_{q,p}(\mathcal{\mathfrak{sl}}(2))$-modules,
the action of the deformed Virasoro algebra can be defined \cite{kon9709}.

The deformed Virasoro algebra is introduced in \cite{LP9412,FR95,SKAO9507}.
It contains two parameters, usually denoted by $q$ and $t$ (and $p=q/t$).
But in order to avoid confusion with those of $U_{q,p}(\widehat{\mathfrak{sl}}(2))$,
we denote them by $\tilde{q}$ and $\tilde{t}$ (and $\tilde{p} = \tilde{q}/\tilde{t}$) in this paper.

The generators of $q$-Virasoro algebra are combined into the generating operator as
\be
\mathcal{T}(z) = \sum_{n \in \mathbb{Z}} \mathcal{T}_n z^{-n}.
\ee
The defining relations of the $q$-Virasoro algebra are given by
\be
f(z_2/z_1) \mathcal{T}(z_1) \mathcal{T}(z_2) - f(z_1/z_2) \mathcal{T}(z_2) \mathcal{T}(z_1)
= \frac{(1-\tilde{q})(1 - \tilde{t}^{-1})}{(1-\tilde{p})}
\bigl[ \delta( \tilde{p} z_1/z_2) - \delta(\tilde{p}^{-1} z_1/z_2) \bigr],
\ee
where
\be
f(z) = \exp\left( \sum_{n>0} \frac{1}{n} \frac{(1-\tilde{q}^n)(1-\tilde{t}^{-n})}{(1+\tilde{p}^n)}
z^n \right).
\ee
Let
\be
\mathcal{T}(z) = \widehat{\Lambda}_1(z) + \widehat{\Lambda}_2(z),
\ee
with
\be
\widehat{\Lambda}_1(z) = q^{P+1} (p^*)^{-(1/2) h} : \exp\left( \sum_{m \neq 0}
\frac{(1-p^{*m})}{[2m]_q} \alpha_m z^{-m} \right):,
\ee
\be
\widehat{\Lambda}_2(z) = q^{-P-1} (p^*)^{(1/2) h}
: \exp\left( - \sum_{m \neq 0} \frac{(1-p^{*m})}{[2m]_q} \alpha_m (q^2 z)^{-m} \right):.
\ee
This operator defines the action of the $q$-Virasoro algebra on the level $1$ module
of $U_{q,p}(\widehat{\mathfrak{sl}}(2))$. The parameters of $q$-Virasoro algebra $\tilde{q}$,
$\tilde{t}$ are respectively identified with those of the elliptic algebra as follows:
\bel{qprel}
\tilde{q} = p, \qq
\tilde{t} = p^* = p q^{-2}.
\ee
Note that
$\tilde{p} = \tilde{q}/\tilde{p} = q^2$.
The parameter $\beta$ is defined by $\tilde{t} = \tilde{q}^{\beta}$. Then we have
$p^* = p^{\beta}$.
If we write
\be
p = q^{2M}, \qq
p^* = q^{2(M-1)},
\ee
then we can read off the value of the parameter $\beta$:
\bel{betaM}
\beta = \frac{M-1}{M}.
\ee
The deformation parameters $\tilde{q}$ and
$\tilde{t}$ is related to the gauge theory parameters as $\tilde{q} = \ex^{R \epsilon_2}$,
$\tilde{t} = \ex^{- R \epsilon_1}$. Therefore, \eqref{qprel} leads to \eqref{dict}.

Let us introduce the following currents:
\be
\tilde{e}(z) = e(z) z^{- (M-1)^{-1} P }, \qq
\tilde{f}(z) = f(z) z^{M^{-1}(P+h) }.
\ee
These are the screening currents of the $q$-Virasoro algebra as we have stressed in the introduction:
\be
[ \mathcal{T}(z_1), \tilde{e}(z_2) ] = (1-p)(1-p^*) \frac{\de_{p^*}}{\de_{p^*} z_2}
\Bigl[ \delta(q z_1/z_2) z_2 \mathcal{O}_1(z_2) \Bigr],
\ee
\be
[ \mathcal{T}(z_2), \tilde{f}(z_2)] = (1-p) (1-p^*) \frac{\de_{p}}{\de_p z_2}
\Bigl[ \delta(q^2z_1/z_2) z_2 \mathcal{O}_2(z_2) \Bigr].
\ee
Here 
\be
\frac{\de_{\xi}}{\de_{\xi} z} f(z) = \frac{f(z) - f(\xi z)}{(1-\xi)z}, \qq
(\xi = p, p^*).
\ee
The operators $\mathcal{O}_1(z)$ and $\mathcal{O}_2(z)$ are given by
\be
\begin{split}
\mathcal{O}_1(z) &= p^{-1} \ex^{-2Q+2Q_h}
q^{P+1} (p^*)^{-(1/2)h} z^{-(M-1)^{-1} P + h} \cr
& \times : \exp\left( - \sum_{m \neq 0} \frac{(1+p^m)}{[2m]_q} \alpha_m (qz)^{-m} \right):,
\end{split}
\ee
\be
\begin{split}
\mathcal{O}_2(z) &= (p^*)^{-1} \ex^{-2Q_h} q^{-(P+1)} (p^*)^{(1/2)h}
z^{M^{-1}(P+h)-h} \cr
& \times : \exp\left( \sum_{m \neq 0} \frac{(1-p^{*m})(1+p^{*m})}{[2m]_q (1-p^m)}
\alpha_m (q^{-2} z)^{-m} \right):.
\end{split}
\ee

\section{Root of unity limit}

We have obtained the dictionary between the parameters of the deformed Virasoro algebra and
those of $U_{q,p}(\widehat{\mathfrak{sl}}(2))$. 
We can take the same root of unity limit of
the parameters as was done in \cite{IOY2,IOY1408}.
 
Let us consider the following $r$-th root of unity limit of the level-$1$ representations of 
$U_{q,p}(\widehat{\mathfrak{sl}}(2))$.
\bel{prl}
p \rightarrow \omega^{\ell}, \qq
p^* \rightarrow \omega^{\ell}, \qq
q^2 \rightarrow 1, 
\ee
where $\omega$ is the primitive $r$-th root of unity $\omega = \exp(2\pi \im/r)$,
and $\ell$ is an integer such that $\omega^{\ell} \neq 1$.

A branch of $\log p$ and $\log p^*$ are chosen as
\be
\log p = 2 \pi \im \left( k_1+ \frac{\ell}{r} \right) - \frac{1}{\sqrt{\beta}} \, R,
\qq (k_1 \in \mathbb{Z}),
\ee
\be
\log p^* = 2 \pi \im \left( k_2 + \frac{\ell}{r} \right) - \sqrt{\beta}\, R,
\qq (k_2 \in \mathbb{Z}),
\ee
and the root of unity limit is meant by the limit of $R \rightarrow 0$.
For simplicity, we assume $k_1 \neq k_2$.

The parameter $\beta$ is restricted to the value
\be
\beta = \frac{k_2 + \ell/r}{k_1 + \ell/r} = \frac{r k_2 + \ell}{r k_1 + \ell}.
\ee
from the consistency of the relation
\be
\beta = \frac{\log p^*}{\log p}.
\ee
For later convenience, let
\bel{mk12}
m:= \frac{r k_1 + \ell}{k_1 - k_2}.
\ee
Then
\bel{betam}
m - r = \frac{r k_2 + \ell}{k_1- k_2}, \qq
\beta = \frac{m-r}{m}.
\ee

By repeating the analysis of \cite{IOY1408}, we see that the level $1$ $U_{q,p}(\widehat{\mathfrak{sl}}(2))$
goes to the tensor product of an algebra  generated by $\{ P, \ex^{\pm 2Q} \}$, the
$\mathbb{Z}_r$-parafermions and a free boson. The parafermions and the boson 
are fields on the $w$-plane where $w=z^r$.

The $\mathbb{Z}_r$-parafermions and the boson in a backgound charge $Q_E/\sqrt{r}$ are described by a conformal field theory (CFT) with the central charge 
\be
c = c_{\mathrm{parafermion}} + c_{\mathrm{boson}},
\ee
with 
\be
c_{\mathrm{parafermion}} = \frac{2(r-1)}{r+2}, \qq
c_{\mathrm{boson}}
=1  -6  \left( \frac{Q_E}{\sqrt{r}} \right)^2, \qq
Q_E = \sqrt{\beta} - \frac{1}{\sqrt{\beta}}.
\ee
Then
\bel{cpL}
\begin{split}
c
&= \frac{2(r-1)}{r+2} +1 - \frac{6r}{m(m-r)} \cr
&= \frac{3r}{r+2} + \frac{3(m-2-r)}{m-r}
- \frac{3 (m-2)}{m},
\end{split}
\ee
which is the central charge of the coset CFT:
\bel{cosetCFTr}
\frac{\widehat{\mathfrak{sl}}(2)_r \oplus \widehat{\mathfrak{sl}}(2)_{m-2-r}}
{\widehat{\mathfrak{sl}}(2)_{m-2}}.
\ee
We remark that by setting $b = \im \sqrt{\beta}$, this central charge \eqref{cpL} can be expressed as
\be
c = \frac{3r}{r+2} + \frac{6}{r} ( b +1/b)^2.
\ee
Hence the coset \eqref{cosetCFTr} may be described by the $r$-th para-Liouville theory
\cite{LNW9206,NT1106}.

\section*{Acknowledgments}
We thank Hidetoshi Awata and Hitoshi Konno for valuable discussions.
We also thank Yaroslav Pugai for useful comments.
This work was supported by JSPS KAKENHI Grant Number 15K05059.
Support from JSPS/RFBR bilateral collaborations 
``Faces of matrix models in quantum field theory
and statistical mechanics'' (H.~I. and R.~Y.)  and
``Exploration of Quantum Geometry
via Symmetry and Duality'' (T.~O.) is gratefully appreciated.


\appendix

\section{Frenkel-Kac construction for $\widehat{\mathfrak{sl}}(2)_1$}
\label{FKC}

We briefly review the Frenkel-Kac construction \cite{FK80} for the case of the level $1$ $\widehat{\mathfrak{sl}}(2)$ current algebra.

Let $\phi(z)$ be the free massless chiral boson on a circle of radius $R$:
\bel{phiz}
\phi(z) = \hat{q} - \im \hat{p} \log z + \im \sum_{n \neq 0} \frac{1}{n} \hat{\alpha}_n z^{-n},
\ee
where
\be
[ \hat{q}, \hat{p} ] = \im, \qq [ \hat{\alpha}_m, \hat{\alpha}_n ] = m \delta_{m+n,0}.
\ee
Note that $\langle \phi(z_1) \phi(z_2) \rangle = - \log(z_1-z_2)$. Let $| 0 \rangle$ be the Fock vacuum defined by
\be
\hat{p} \, | 0 \rangle = 0, \qq
\hat{\alpha}_n | 0 \rangle = 0, \qq
(n > 0).
\ee
Since the boson is compactified on the circle with radius $R$, the eigenvalues of the momentum operator $\hat{p}$
must be $n/R$ $(n \in \mathbb{Z})$. Let us denote the corresponding momentum eigenstates by
\be
|n; R \rangle = \ex^{\im (n/R) \hat{q} } | 0 \rangle, \qq \hat{p}\,  | n; R \rangle
= \frac{n}{R} \, | n; R \rangle.
\ee
The compactified boson Fock space $\mathcal{F}_R$ is obtained from the Fock vacuum $| 0 \rangle$
by acting the creation operators $\hat{\alpha}_{-m}$ $(m>0)$ and $\ex^{\im (n/R) \hat{q}}$
($n \in \mathbb{Z}$).
On this Fock space the action of the position operator $\hat{q}$ is allowed only through the form of
$\ex^{\im (n/R) \hat{q}}$ $(n \in \mathbb{Z})$. Hence the vertex operators
\be
V_{u}(z) = : \ex^{\im u \phi(z) } :
= \ex^{\im u \hat{q}} z^{u \hat{p}} 
\exp\left( u \sum_{n>0} \frac{1}{n} \hat{\alpha}_{-n} z^n \right)
\exp\left( - u \sum_{n>0} \frac{1}{n} \hat{\alpha}_n z^{-n} \right)
\ee
with $R\, u \in \mathbb{Z}$ can act on the Fock space.
Let
\be
J^{\pm}(z) = V_{\pm \sqrt{2}}(z) = : \ex^{\pm \im \sqrt{2} \phi(z)}: , \qq 
J^3(z) = \frac{\im}{\sqrt{2}} \partial \phi(z),
\ee
with mode expansion
\be
J^{a}(z) = \sum_{m \in \mathbb{Z}} J^a_m z^{-m-1}, \qq a=\pm, 3.
\ee
It is well-known that $J^a_m$ realize the affine Lie algebra $\widehat{\mathfrak{sl}}(2)$ with level $1$.
In order to make the Fock space $\mathcal{F}_R$ be a $\widehat{\mathfrak{sl}}(2)_1$-module, the compactification
radius $R$ must be an integer multiple of $1/\sqrt{2}$. 

In particular, at $R=\sqrt{2}$
(the self-dual radius), $J^{\pm}_m$ shifts the momentum $\hat{p}=n/\sqrt{2}$ to $\hat{p} = (n \pm 2)/\sqrt{2}$.
Therefore,
the boson Fock space $\mathcal{F}_{R=\sqrt{2}}$ decomposes into two 
irreducible $\widehat{\mathfrak{sl}}(2)_1$-modules according to the values of $(-1)^{\sqrt{2} \hat{p}}$:
\be
\mathcal{F}_{\sqrt{2}} = L(\Lambda_0) \oplus L(\Lambda_1),
\ee
where
\be
L(\Lambda_0) = \{ v \in \mathcal{F}_{\sqrt{2}} \, | \, (-1)^{\sqrt{2} \hat{p}} \, v = (+1) v \},
\qq
L(\Lambda_1) = \{ v \in \mathcal{F}_{\sqrt{2}} \, | \, (-1)^{\sqrt{2} \hat{p}} \, v = (-1) v \}.
\ee
Here $\Lambda_0$ and $\Lambda_1$ are the fundamental weights of  
the affine Lie algebra $\widehat{\mathfrak{sl}}(2)$ and
$L(\Lambda_i)$ are the integrable highest-weight module with highest weight $\Lambda_i$.
The highest-weight vector of the basic module $L(\Lambda_0)$ and that of the defining module
$L(\Lambda_1)$ are respectively given by
\be
|\Lambda_0 \rangle = | 0 ; \sqrt{2} \, \rangle = | 0 \rangle \in L(\Lambda_0), \qq
|\Lambda_1 \rangle = | 1 ; \sqrt{2} \, \rangle
= \ex^{(\im/\sqrt{2}) \hat{q} }| 0 \rangle \in L(\Lambda_1).
\ee

\vspace{3mm}

\noindent
\textbf{Remark}. The boson $\Phi_0(z)$ \eqref{Phiz} is related to the canonically normalized
boson $\phi(z)$ \eqref{phiz} as $\Phi_0(z) = \im \sqrt{2} \, \phi(z)$.
In particular, $h$ is identified with $\sqrt{2}\,  \hat{p}$. Hence at the self-dual radius, the eigenvalues of $h$
must be integers.




\begin{thebibliography}{99}

\bibitem{AGT0906}
L.~F. Alday, D.~Gaiotto, and Y.~Tachikawa,
``{Liouville Correlation Functions from Four-dimensional Gauge Theories},''
Lett. Math. Phys. {\bf 91}, 167--197 (2010) [arXiv:0906.3219 [hep-th]].

\bibitem{wyl0907}
N.~Wyllard,
``{$A_{N-1}$ conformal Toda field theory correlation functions from conformal
  $\mathcal{N}=2$ $SU(N)$ quiver gauge theories},''
JHEP {\bf 0911}, 002 (2009) [arXiv:0907.2189 [hep-th]].

\bibitem{wyl1109}
N.~Wyllard, ``Coset conformal blocks and $\mathcal{N}=2$ gauge theories,''
  arXiv:1109.4264 [hep-th].

\bibitem{AT1110}
M.~N.~Alfimov and G.~M.~Tarnopolsky, ``Parafermionic Liouville field theory and
  instantons on ALE spaces,'' JHEP {\bf 1202}, 036 (2012) [arXiv:1110.5628
  [hep-th]].

\bibitem{BFL1011}
M.~A. Bershtein, V.~A. Fateev, and A.~V. Litvinov,
``{Parafermionic polynomials, Selberg integrals and three-point correlation
  function in parafermionic Liouville field theory},''
Nucl. Phys. {\bf B847}, 413--459 (2011) [arXiv:1011.4090 [hep-th]].

\bibitem{BF1105}
V.~Belavin and B.~Feigin, ``Super Liouville conformal blocks from
  $\mathcal{N}=2$ $SU(2)$ quiver gauge theories,'' JHEP {\bf 1107}, 079 (2011)
  [arXiv:1105.5800 [hep-th]].

\bibitem{BBB1106}
A.~Belavin, V.~Belavin and M.~Bershtein, ``Instantons and 2d Superconformal
  field theory,'' JHEP {\bf 1109}, 117 (2011) [arXiv:1106.4001 [hep-th]].

\bibitem{BMT}
G.~Bonelli, K.~Maruyoshi and A.~Tanzini, ``Instantons on ALE spaces and super
  Liouville conformal field theories,'' JHEP {\bf 1108}, 056 (2011)
  [arXiv:1106.2505 [hep-th]]; ``Gauge Theories on ALE Space and Super Liouville
  Correlation Functions,'' Lett.\ Math.\ Phys.\ {\bf 101}, 103-124 (2012)
  [arXiv:1107.4609 [hep-th]].

\bibitem{I1110}
Y.~Ito, ``Ramond sector of super Liouville theory from instantons on an ALE
  space,'' Nucl.\ Phys.\ B {\bf 861}, 387-402 (2012) [arXiv:1110.2176
  [hep-th]].

\bibitem{BBFLT1111}
A.~A.~Belavin, M.~A.~Bershtein, B.~L.~Feigin, A.~V.~Litvinov and
  G.~M.~Tarnopolsky, ``Instanton moduli spaces and bases in coset conformal
  field theory,'' Commun.\ Math.\ Phys.\ {\bf 319}, 269-301 (2013)
  [arXiv:1111.2803 [hep-th]].

\bibitem{BBT1211}
A.~A.~Belavin, M.~A.~Bershtein and G.~M.~Tarnopolsky, ``Bases in coset
  conformal field theory from AGT correspondence and Macdonald polynomials at
  the roots of unity,'' JHEP {\bf 1303}, 019 (2013) [arXiv:1211.2788 [hep-th]].

\bibitem{ABT1306}
M.~N.~Alfimov, A.~A.~Belavin and G.~M.~Tarnopolsky, ``Coset conformal field
  theory and instanton counting on $\mathbb{C}^2/\mathbb{Z}_p$,'' JHEP {\bf
  1308}, 134 (2013) [arXiv:1306.3938 [hep-th]].

\bibitem{IOY2}
H.~Itoyama, T.~Oota and R.~Yoshioka, ``2d-4d Connection between $q$-Virasoro/W
  Block at Root of Unity Limit and Instanton Partition Function on ALE Space,''
  Nucl.\ Phys.\ B {\bf 877}, 506-537 (2013) [arXiv:1308.2068 [hep-th]];
  ``q-Virasoro algebra at root of unity limit and 2d-4d connection,'' J.\
  Phys.\ Conf.\ Ser.\ {\bf 474}, 012022 (2013).

\bibitem{IOY1408}
H.~Itoyama, T.~Oota and R.~Yoshioka, ``$q$-Virasoro/W Algebra at Root of Unity
  and Parafermions,'' Nucl.\ Phys.\ B {\bf 889}, 25-35 (2014) [arXiv:1408.4216
  [hep-th]].

\bibitem{BPSS1312}
  U.~Bruzzo, M.~Pedrini, F.~Sala and R.~J.~Szabo,
  ``Framed sheaves on root stacks and supersymmetric gauge theories on ALE spaces,''
  Adv.\ Math.\  {\bf 288}, 1175-1308 (2016)
  [arXiv:1312.5554 [math.AG]]; \\
    M.~Pedrini, F.~Sala and R.~J.~Szabo,
  ``AGT relations for abelian quiver gauge theories on ALE spaces,''
  J.\ Geom.\ Phys.\  {\bf 103}, 43-89 (2016)
  [arXiv:1405.6992 [math.RT]];\\
    U.~Bruzzo, F.~Sala and R.~J.~Szabo,
  ``${\mathcal{N} = 2}$ Quiver Gauge Theories on A-type ALE Spaces,''
  Lett.\ Math.\ Phys.\  {\bf 105}, 401-445 (2015)
  [arXiv:1410.2742 [hep-th]].


\bibitem{NT1106}
T.~Nishioka and Y.~Tachikawa,
``{Central charges of para-Liouville and Toda theories from M5-branes},''
Phys. Rev. {\bf D84}, 046009 (2011) [arXiv:1106.1172 [hep-th]].

\bibitem{AY0910}
H.~Awata and Y.~Yamada,
``{Five-dimensional AGT conjecture and the deformed Virasoro algebra},''
JHEP {\bf 1001}, 125 (2010) [arXiv:0910.4431 [hep-th]]; 
``{Five-Dimensional AGT Relation and the Deformed $\beta$-Ensemble},''
Prog. Theor. Phys. {\bf 124}, 227--262 (2010) [arXiv:1004.5122 [hep-th]].

\bibitem{NPP1303}
F.~Nieri, S.~Pasquetti and F.~Passerini, ``3d and 5d Gauge Theory Partition
  Functions as $q$-deformed CFT Correlators,'' Lett.\ Math.\ Phys.\ {\bf 105},
  109-148 (2015) [arXiv:1303.2626 [hep-th]]; \\
F.~Nieri, S.~Pasquetti, F.~Passerini and A.~Torrielli, ``5D partition
  functions, $q$-Virasoro systems and integrable spin-chains'' JHEP {\bf 1412},
  040 (2014) [arXiv:1312.1294 [hep-th]].

\bibitem{Orl1310}
D.~Orlando, ``A stringy perspective on the quantum integrable model/gauge
  correspondence,'' arXiv:1310.0031 [hep-th].

\bibitem{tan1301}
M.-C. Tan,
``{M-theoretic derivations of 4d-2d dualities: from a geometric Langlands
  duality for surfaces, to the AGT correspondence, to integrable systems},''
JHEP {\bf 1307}, 171 (2013) [arXiv:1301.1977 [hep-th]].

\bibitem{AHKS1309}
M.~Aganagic, N.~Haouzi, C.~Koz\c{c}az, and S.~Shakirov,
``{Gauge/Liouville Triality},''
arXiv:1309.1687 [hep-th].

\bibitem{yos1512}
R.~Yoshioka, ``The integral representation of solutions of KZ equation and a
  modification by $\mathcal{K}$ operator insertion,'' arXiv:1512.01084
  [hep-th].

\bibitem{KP1512}
T.~Kimura and V.~Pestun,
``{Quiver W-algebras},''
arXiv:1512.08533 [hep-th].

\bibitem{sai1301}
Y.~Saito,
``{Elliptic Ding-Iohara Algebra and the Free Field Realization of the Elliptic
  Macdonald Operator},''
Publ. Res. Inst. Math. Sci. {\bf 50}, 411--455 (2014) [arXiv:1301.4912
  [math.QA]]; 
``{Commutative Families of the Elliptic Macdonald Operator},''
SIGMA {\bf 10}, 021 [17 pages] (2014) [arXiv:1305.7097 [math.QA]].

\bibitem{Tan1309}
M.-C. Tan,
``{An M-theoretic derivation of a 5d and 6d AGT correspondence, and
  relativistic and elliptized integrable systems},''
JHEP {\bf 1312}, 031 (2013) [arXiv:1309.4775 [hep-th]]; 
``{Higher AGT Correspondences, W-algebras, and Higher Quantum Geometric
  Langlands Duality from M-Theory},''
arXiv:1607.08330 [hep-th].

\bibitem{IKY1511}
A.~Iqbal, C.~Koz\c{c}az, and S.-T. Yau,
``{Elliptic Virasoro Conformal Blocks},''
arXiv:1511.00458 [hep-th].

\bibitem{nie1511}
F.~Nieri,
``{An elliptic Virasoro symmetry in 6d},''
arXiv:1511.00574 [hep-th].

\bibitem{KP1608}
T.~Kimura and V.~Pestun,
``{Quiver elliptic W-algebras},''
arXiv:1608.04651 [hep-th].

\bibitem{LP9412}
S.~Lukyanov and Ya.~Pugai, ``Bosonization of ZF algebras: Direction toward
  deformed Virasoro algebra,'' J.\ Exp.\ Theor.\ Phys.\ {\bf 82}, 1021-1045
  (1996) [Zh.\ Eksp.\ Teor.\ Fiz.\ {\bf 109}, 1900-1947 (1996)]
  [arXiv:hep-th/9412128].

\bibitem{FR95}
E.~Frenkel and N.~Reshetikhin, ``Quantum Affine Algebras and Deformations of
  the Virasoso and $\mathcal{W}$-Algebras,'' Commun.\ Math.\ Phys.\ {\bf 178},
  237-264 (1996) [arXiv:q-alg/9505025].

\bibitem{SKAO9507}
J.~Shiraishi, H.~Kubo, H.~Awata and S.~Odake, ``A quantum deformation of the
  Virasoro algebra and the Macdonald symmetric functions,'' Lett.\ Math.\
  Phys.\ {\bf 38}, 33-51 (1996) [arXiv:q-alg/9507034].

\bibitem{FF9508}
B.~Feigin and E.~Frenkel, ``Quantum $\mathcal{W}$-Algebras and Elliptic
  Algebras,'' Commun.\ Math.\ Phys.\ {\bf 178}, 653-678 (1996)
  [arXiv:q-alg/9508009].

\bibitem{AKOS9508}
H.~Awata, H.~Kubo, S.~Odake and J.~Shiraishi, ``Quantum $\mathcal{W}_N$
  Algebras and Macdonald Polynomials,'' Commun.\ Math.\ Phys.\ {\bf 179},
  401-416 (1996) [arXiv:q-alg/9508011].

\bibitem{AKMOS9604}
H.~Awata, H.~Kubo, Y.~Morita, S.~Odake and J.~Shiraishi, ``Vertex Operators of
  the $q$-Virasoro Algebra; Defining Relations, Adjoint Actions and Four Point
  Functions,'' Lett.\ Math.\ Phys.\ {\bf 41}, 65-78 (1997)
  [arXiv:q-alg/9604023].

\bibitem{kad}
A.~A.~Kadeishvili, ``Vertex operators for deformed Virasoro algebra,'' JETP
  Lett.\ {\bf 63}, 917-923 (1996) [Pisma Zh.\ Eksp.\ Teor.\ Fiz.\ {\bf 63},
  876-881 (1996)] [arXiv:hep-th/9604153].

\bibitem{JS97}
M.~Jimbo and J.~Shiraishi, ``A Coset-Type Construction for the Deformed
  Virasoro Algebra,'' Lett.\ Math.\ Phys.\ {\bf 43}, 173-185 (1998)
  [arXiv:q-alg/9709037].

\bibitem{IOY1602}
H.~Itoyama, T.~Oota, and R.~Yoshioka,
``{$q$-vertex operator from 5D Nekrasov function},''
J. Phys. {\bf A49}, 345201 (2016) [arXiv:1602.01209 [hep-th]].

\bibitem{DI9608a}
J.~Ding and K.~Iohara,
``{Generalization of Drinfeld Quantum Affine Algebras},''
Lett. Math. Phys. {\bf 41}, 181--193 (1997) [arXiv:q-alg/9608002].

\bibitem{mik07}
K.~Miki,
``{A $(q, \gamma)$ analog of the $W_{1+\infty}$ algebra},''
J. Math. Phys. {\bf 48}, 123520 (2007).

\bibitem{kon9709}
H.~Konno,
``{An Elliptic Algebra $U_{q,p}(\widehat{sl_2})$ and the Fusion RSOS Model},''
Commun. Math. Phys. {\bf 195}, 373--403 (1998) [arXiv:q-alg/9709013].

\bibitem{JKOS9802}
M.~Jimbo, H.~Konno, S.~Odake, and J.~Shiraishi,
``{Elliptic Algebra $U_{q,p}(\widehat{\mathfrak{sl}}_2)$: Drinfeld Currents and
  Vertex Operators},''
Commun. Math. Phys. {\bf 199}, 605--647 (1999) [arXiv:math/9802002 [math.QA]].

\bibitem{KK0210}
T.~Kojima and H.~Konno,
``{The Elliptic Algebra $U_{q,p}(\widehat{\mathfrak{sl}}_N)$ and the Drinfeld
  Realization of the Elliptic Quantum Group
  $\mathcal{B}_{q,\lambda}(\widehat{\mathfrak{sl}}_N)$},''
Commun. Math. Phys. {\bf 239}, 405--447 (2003) [arXiv:math/0210383 [math.QA]].

\bibitem{KK0307}
T.~Kojima and H.~Konno,
``{The elliptic algebra $U_{q,p}(\widehat{\mathfrak{sl}}_N)$ and the
  deformation of $W_N$ algebra},''
J. Phys. A: Math. Gen. {\bf 37}, 371--383 (2004) [arXiv:math/0307244
  [math.QA]].

\bibitem{KKW0504}
T.~Kojima, H.~Konno, and R.~Weston,
``{The vertex-face correspondence and correlation functions of the fusion
  eight-vertex model: I: The general formalism},''
Nucl. Phys. {\bf B720}, 348--398 (2005) [arXiv:math/0504433 [math.QA]].

\bibitem{kon0802}
H.~Konno,
``{Elliptic quantum group $U_{q,p}(\widehat{\mathfrak{sl}}_2)$ and vertex
  operators},''
J. Phys. {\bf A41}, 194012 (2008) [arXiv:0802.3630 [math.QA]].

\bibitem{kon0803}
H.~Konno,
``{Elliptic quantum group $U_{q,p}(\widehat{\mathfrak{sl}}_2)$, Hopf algebroid
  structure and elliptic hypergeometric series},''
J. Geom. Phys. {\bf 59}, 1485--1511 (2009) [arXiv:0803.2292 [math.QA]].

\bibitem{kon1603}
H.~Konno,
``{Elliptic Quantum Groups $U_{q,p}(\widehat{\mathfrak{gl}}_N)$ and
  $E_{q,p}(\widehat{\mathfrak{gl}}_N)$},''
arXiv:1603.04129 [math.QA].

\bibitem{AFHKSY}
H.~Awata, B.~Feigin, A.~Hoshimo, M.~Kanai, J.~Shiraishi and S.~Yanagida,
  ``Notes on Ding-Iohara algebra and AGT conjecture,'' RIMS K\^{o}ky\^{u}roku
  {\bf 1765}, 12-32 (2011) [arXiv:1106.4088 [math-ph]].

\bibitem{KMZ1308}
S.~Kanno, Y.~Matsuo and H.~Zhang, ``Extended Conformal Symmetry and Recursion
  Formulae for Nekrasov Partition Function,'' JHEP {\bf 1308}, 028 (2013)
  [arXiv:1306.1523 [hep-th]]; \\
Y.~Matsuo, C.~Rim and H.~Zhang, ``Construction of Gaiotto states with
  fundamental multiplets through degenerate DAHA,'' JHEP {\bf 1409}, 028 (2014)
  [arXiv:1405.3141 [hep-th]]; \\
J.-E. Bourgine, Y.~Matsuo, and H.~Zhang,
``{Holomorphic field realization of SH$^{c}$ and quantum geometry of quiver
  gauge theories},''
JHEP {\bf 1604}, 167 (2016) [arXiv:1512.02492 [hep-th]].

\bibitem{spo1409}
L.~Spodyneiko,
  ``AGT correspondence: Ding–Iohara algebra at roots of unity and Lepowsky–Wilson construction,''
  J.\ Phys.\ A {\bf 48}, 275404 (2015)
  [arXiv:1409.3465 [hep-th]].

\bibitem{OAF1512}
Y.~Ohkubo, H.~Awata, and H.~Fujino,
``{Crystallization of deformed Virasoro algebra, Ding-Iohara-Miki algebra and
  5D AGT correspondence},''
arXiv:1512.08016 [math-ph].

\bibitem{AKMMMMOZ1604}
H.~Awata, H.~Kanno, T.~Matsumoto, A.~Mironov, A.~Morozov, And.~Morozov,
  Y.~Ohkubo, and Y.~Zenkevich,
``{Explicit examples of DIM constraints for network matrix models},''
JHEP {\bf 1607}, 103 (2016) [arXiv:1604.08366 [hep-th]].

\bibitem{AKMMMOZ1608}
H.~Awata, H.~Kanno, A.~Mironov, A.~Morozov, And.~Morozov, Y.~Ohkubo, and
  Y.~Zenkevich,
``{Toric Calabi-Yau threefolds as quantum integrable systems. R-matrix and RTT
  relations},''
JHEP {\bf 1610}, 047 (2016)  
[arXiv:1608.05351 [hep-th]]; 
``{Anomaly in RTT relation for DIM algebra and network matrix models},''
Nucl. Phys. {\bf B918}, 358--385 (2017) [arXiv:1611.07304 [hep-th]]; 
``{$(q,t)$-KZ equation for Ding-Iohara-Miki algebra},''
arXiv:1703.06084 [hep-th].

\bibitem{MMZ1603a}
A.~Mironov, A.~Morozov, and Y.~Zenkevich,
``{Spectral duality in elliptic systems, six-dimensional gauge theories and
  topological strings},''
JHEP {\bf 1605}, 121 (2016) [arXiv:1603.00304 [hep-th]]; 
``{Ding-Iohara-Miki symmetry of network matrix models},''
Phys. Lett. {\bf B762}, 196--208 (2016) [arXiv:1603.05467 [hep-th]].

\bibitem{BFMZZ1606}
J.-E. Bourgine, M.~Fukuda, Y.~Matsuo, H.~Zhang, and R.-D. Zhu,
``{Coherent states in quantum $\mathcal{W}_{1+\infty}$ algebra and qq-character
  for 5d super Yang-Mills},''
PTEP {\bf 2016}, 123B05 (2016) [arXiv:1606.08020 [hep-th]]; \\
J.-E. Bourgine, M.~Fukuda, K.~Harada, Y.~Matsuo, and R.-D. Zhu,
``{$(p,q)$-webs of DIM representations, 5d $\mathcal{N}=1$ instanton partition
  functions and qq-characters},''
arXiv:1703.10759 [hep-th]; \\
M.~Fukuda, K.~Harada, Y.~Matsuo and R.-D.~Zhu,
  ``Maulik-Okounkov's R-matrix from Ding-Iohara-Miki algebra,''
  arXiv:1705.02941 [hep-th].


\bibitem{bax72}
R.~J. Baxter,
``{Partition Function of the Eight-Vertex Lattice Model},''
Annals Phys. {\bf 70}, 193--228 (1972).

\bibitem{bel81}
A.~A. Belavin,
``{Dynamical symmetry of integrable quantum systems},''
Nucl. Phys. {\bf B180}, 189--200 (1981).

\bibitem{ABF84}
G.~E. Andrews, R.~J. Baxter, and P.~J. Forrester,
``{Eight-Vertex SOS Model and Generalized Rogers-Ramanujan-Type Identities},''
J. Statist. Phys. {\bf 35}, 193--266 (1984).

\bibitem{skl82}
E.~K. Sklyanin,
``Some algebraic structures connected with the Yang-Baxter equation,''
Funkt. Anal. Prilozhen. {\bf 16}, 27--34 (1982) 
{[Funct. Anal. Appl. {\bf 16}, 263-270 (1982)]}; \\
I.~V. Cherednik,
``Some finite-dimensional representations of generalized {S}klyanin algebras,''
Funkt. Anal. Prilozhen. {\bf 19}, 89--90 (1985) 
{[Funct. Anal. Appl. {\bf 19}, 77-79 (1985)]}.

\bibitem{FIJKMY9403}
O.~Foda, K.~Iohara, M.~Jimbo, R.~Kedem, T.~Miwa, and H.~Yan,
``{An elliptic quantum algebra for $\widehat{sl}_2$},''
Lett. Math. Phys. {\bf 32}, 259--268 (1994) [arXiv:hep-th/9403094].

\bibitem{JKOS9712}
M.~Jimbo, H.~Konno, S.~Odake, and J.~Shiraishi,
``{Quasi-Hopf twistors for elliptic quantum groups},''
Transform. Groups {\bf 4}, 303--327 (1999) [arXiv:q-alg/9712029].

\bibitem{GN84}
J.-L. Gervais and A.~Neveu,
``{Novel triangle relation and absence of tachyons in Liouville string field
  theory},''
Nucl. Phys. {\bf B238}, 125--141 (1984).

\bibitem{fel9407}
G.~Felder,
``{Conformal field theory and integrable systems associated to elliptic
  curves},''
arXiv:hep-th/9407154.

\bibitem{fel9412}
G.~Felder,
``{Elliptic quantum groups},''
in {\em {Mathematical physics. Proceedings, 11th International Congress, Paris,
  France, July 18-22, 1994}}  211--218 (1994) [arXiv:hep-th/9412207].

\bibitem{bab91}
O.~Babelon,
``{Universal Exchange Algebra for Bloch Waves and Liouville Theory},''
Commun. Math. Phys. {\bf 139}, 619--643 (1991).

\bibitem{BBB96}
O.~Babelon, E.~Billey, and D.~Bernard,
``{A quasi-Hopf algebra interpretation of quantum 3-$j$ and 6-$j$ symbols and
  difference equations.},''
Phys. Lett. {\bf B375}, 89--97 (1996).

\bibitem{ABB9505}
J.~Avan, O.~Babelon, and E.~Billey,
``{The Gervais-Neveu-Felder equation and the quantum Calogero-Moser systems},''
Commun. Math. Phys. {\bf 178}, 281--300 (1996) [arXiv:hep-th/9505091].

\bibitem{dri89}
V.~G. Drinfel'd,
``{Quasi-Hopf algebras},''
Algebra i Analiz {\bf 2}, 149--181 (1989) 
[Leningrad Math. J. {\bf 1}, 1419-1457 (1990)].

\bibitem{EV9708}
P.~Etingof and A.~Varchenko,
``{Solutions of the Quantum Dynamical Yang-Baxer Equation and Dynamical Quantum
  Groups},''
Commun. Math. Phys. {\bf 196}, 591--640 (1998) [arXiv:q-alg/9708015]; 
``{Exchange Dynamical Quantum Groups},''
Commun. Math. Phys. {\bf 205}, 19--52 (1999) [arXiv:math/9801135 [math.QA]].

\bibitem{LP9602}
S.~L.~Lukyanov and Ya.~Pugai, ``Multi-point local height probabilities in the
  integrable RSOS model,'' Nucl.\ Phys.\ B {\bf 473}, 631-658 (1996)
  [arXiv:hep-th/9602074].

\bibitem{JLMP9607}
  M.~Jimbo, M.~Lashkevich, T.~Miwa and Y.~Pugai,
  ``Lukyanov's screening operators for the deformed Virasoro algebra,''
  Phys.\ Lett.\ A {\bf 229}, 285-292 (1997)
  [hep-th/9607177].

\bibitem{pug04}
  Y.~Pugai,
  ``Vacuum expectation values from fusion of vertex operators,''
  JETP Lett.\  {\bf 79}, 457-463 (2004)
  [Pisma Zh.\ Eksp.\ Teor.\ Fiz.\  {\bf 79}, 569-574 (2004)].

\bibitem{FKO1404}
R.~M. Farghly, H.~Konno, and K.~Oshima,
``{Elliptic Algebra $U_{q,p}(\widehat{\mathfrak{g}})$ and Quantum
  $Z$-algebras},''
Algebr. Represent. Theory {\bf 18}, 103--135 (2015) [arXiv:1404.1738
  [math.QA]].

\bibitem{FK80}
I.~B. Frenkel and V.~G. Kac,
``{Basic Representations of Affine Lie Algebras and Dual Resonance Models},''
Invent. math. {\bf 62}, 23--66 (1980).

\bibitem{FJ88}
I.~B. Frenkel and N.~Jing,
``{Vertex representations of quantum affine algebras},''
Proc. Natl. Acad. Sci. USA {\bf 85}, 9373--9377 (1988).

\bibitem{LNW9206}
A.~LeClair, D.~Nemeschansky, and N.~P. Warner,
``{$S$-matrices for perturbed $N=2$ superconformal field theory from quantum
  groups},''
Nucl. Phys. {\bf B390}, 653--680 (1993) [arXiv:hep-th/9206041].




  
\end{thebibliography}
\end{document}